# Moroccan pre-service elementary teachers: attitudes toward STEM education and mobile devices

**Aziz Amaaz[1], Abderrahman Mouradi[2], Moahamed Erradi[1], Ali Allouch[3]**

[1]Computer Science and University Pedagogical Engineering Research Team, Higher Normal School, Abdelmalek Essaadi University, Tetouan, Morocco
[2]Energy, Materials and Computing Physics Research Team, Higher Normal School, Abdelmalek Essaadi University, Tetouan, Morocco
[3]Higher School of Education and Training, Ibn Tofail University, Kenitra, Morocco

| Article Info | ABSTRACT |
|---|---|
| *Article history:*<br><br>Received Jul 30, 2023<br>Revised Mar 3, 2024<br>Accepted Mar 19, 2024<br><br>*Keywords:*<br><br>Attitudes<br>Elementary school<br>Integrated STEM education<br>Mobile devices<br>Pre-service teachers | The purpose of this study was to explore Moroccan pre-service elementary teachers' attitudes toward integrated science, technology, engineering, and mathematics (STEM) education and the use of mobile devices in integrated STEM education. The research sample was selected using convenience sampling. Data were collected from 226 pre-service teachers in the Bachelor of Education Elementary Specialty (BEES) using a 28-item questionnaire. The validity of the items was tested by factor analysis using the extraction method of principal component analysis with varimax rotation. Reliability tests for the different constructs were conducted by calculating Cronbach's alpha. Frequency, mean, standard deviation and Mann-Whitney tests were used to analyze the data. The results revealed that pre-service elementary teachers have generally neutral attitudes toward integrated STEM education, and they also showed that pre-service teachers' attitudes toward integrated STEM education do not depend on gender or grade level. However, these attitudes are dependent on pre-university studies. Pre-service teachers with a scientific background have significantly more positive attitudes toward integrated STEM education than their counterparts with a literary background. Furthermore, the results of this study also revealed that pre-service teachers have positive attitudes toward the use of mobile devices in integrated STEM education, and these attitudes are not dependent on gender, grade level, or pre-university studies.<br><br> |

*Corresponding Author:*

Aziz Amaaz
Computer Science and University Pedagogical Engineering Research Team, Higher Normal School,
Abdelmalek Essaadi University
Sebta Avenue, Mhannech II, 93002 Tetouan, Morocco
E-mail: aamaaz@uae.ac.ma

## 1. INTRODUCTION

The world we live in is constantly changing, and many of the problems we face in this world are interdisciplinary in nature, requiring the integration of a wide range of knowledge and practices from science, technology, engineering, and mathematics (STEM) subjects to solve them [1]. This integration is not an easy task because real-world problems are not compartmentalized in the same way that STEM subjects are taught in school [2]. The complexity of the problems and the difficulty students have in mobilizing their knowledge in separate STEM subjects to solve these problems have led many educational systems to adopt integrated STEM education.

Integrated STEM education is a curricular approach that combines the concepts of STEM in an interdisciplinary teaching approach [1], [3] that links these four fields so that learning becomes connected, focused, meaningful, and relevant to learners [4]. Thus, integrated STEM education creates a learning environment where students can understand the relationships between mathematics, science, engineering, and technology. The goal of the integrated STEM education approach is to change the way science is taught through the introduction of technology and engineering into student activities [3]. This introduction is likely to motivate students to learn science and mathematics and positively change their perceptions of technology and engineering [5], [6].

In spite of the fact that the concept of integrated STEM education has been considered in the United States beginning in the 1990s, how STEM is taught and the relationship between the four fields are still a matter of debate several decades later. Several models are proposed for implementing integrated STEM education. Each model may employ a combination of STEM fields, emphasize one field more than the others, take place in a formal or informal setting, and employ a variety of pedagogical choices [7]. Among the proposed models is one that suggests that integrated STEM education should use engineering and technical design activities as a context for students to make connections between content and practices in STEM fields [8]. In other words, engineering practices and technical design are an integrative component of the content to be learned in other STEM disciplines.

The implementation of integrated STEM education requires a change in instructional practices that cannot succeed without teacher preparation. This preparation depends on several specific elements, including the knowledge, attitudes, and skills needed to implement integrated STEM education [8]–[10]. Notably, teachers' attitudes and beliefs influence their teaching practices, which in turn influence their students' attitudes and beliefs [11], [12]. Thus, teachers can negatively affect their students' classroom activities and attitudes toward STEM, just as they can promote their students' interests and attitudes toward STEM [13], [14]. Teachers with positive attitudes toward STEM tend to enjoy implementing STEM activities in their classrooms, while teachers with negative attitudes tend to avoid STEM-related activities [15], [16]. Stohlmann *et al.* [4] reported that teachers' passion for STEM education influenced their confidence and comfort in adopting this approach.

Studies have shown that teachers are generally aware of the importance of integrated STEM education and believe it should be implemented in K–12 [17], [18]. Overall, teachers believe that the interdisciplinary nature of integrated STEM education and incorporating design activities are beneficial to students' learning and their futures [8], [17]. Teachers also believe that integrated STEM instruction motivates students and increases student engagement [8], [19].

The teaching of integrated STEM education is sometimes challenged not only by a lack of teaching materials, laboratories, and learner motivation but also by ineffective didactic choices implemented by teachers [20], [21]. The use of mobile devices has the potential to mitigate the impacts of these challenges by providing teachers with the ability to implement multiple pedagogical and didactic choices (e.g., educational games, quizzes, group work, individualized learning, situational learning, and out-of-class learning) [22], [23]. Mobile devices also enable anytime, anywhere access to learning and assessment materials [24], can have a positive impact on learners' motivation to learn [25], and allow visualization of scientific experiments, which can improve learners' understanding of scientific, mathematical, and technological concepts [26].

Although mobile devices can provide benefits to integrated STEM education, their availability does not guarantee their use in this education; therefore, it is important to assess teachers' readiness to use mobile devices [23], [27] by examining their attitudes toward this technology. Teachers' attitudes toward mobile devices are a key factor in understanding their motivation to use this technology in integrated STEM education. Examining these attitudes will help policymakers take initiatives to adopt mobile devices in integrated STEM education and to set up the appropriate infrastructure.

The concept of attitude has its origins in social psychology, where it has been a primary concern throughout most of the last century [28], [29]. Researchers have been interested in attitude to the extent that it can predict and affect the actions and behaviors of individuals. In social psychology, attitude refers to an evaluative judgment that integrates and summarizes a person's cognitive and affective reactions to an object/behavior [28], [30]. Thus, a person may have a negative, neutral, or positive attitude toward the object or behavior.

Attitudes are not directly observable; they are often inferred from indicators of three components: affective, behavioral, and cognitive [29]–[31]. The affective component consists of a person's sentiments and emotions toward the object of the attitude [30], [31]. The behavioral component includes the ways in which a person acts or behaves in relation to the object of their attitude [30], [31]. The cognitive component includes a person's knowledge and beliefs about the attitude object [31]. There is no consensus among researchers that the cognitive, affective, and behavioral components must be present for an attitude to emerge. An attitude may be formed primarily or exclusively by one of these three components, depending on the nature of the object of the attitude and the relationship the individual establishes with that object [31].

To examine teachers' attitudes toward the use of mobile devices and the factors that motivate their intentions to use or not use this technology, several studies have been conducted using models developed to explain technology acceptance. The technology acceptance model (TAM) is one of the most widely used models to explain user behavior across a wide range of computing technologies and users [32]. According to the TAM model, perceived usefulness and perceived ease of use are important determinants of technology acceptance behavior [33]–[35]. Perceived usefulness refers to an individual's belief in the technology's ability to improve their performance [33]. Conversely, ease of use refers to an individual's expectation that using the technology will require minimal effort [33]. In addition to the two core TAM constructs (perceived usefulness and perceived ease of use), studies [36], [37] have suggested adding perceived enjoyment as a strong construct to improve the TAM model's ability to explain people's technology adoption intentions. Intention to adopt technology is defined as the degree of willingness to use the technology on an ongoing basis [33], whereas perceived enjoyment refers to the extent to which the activity of using the technology is perceived as enjoyable in and of itself, regardless of any expected performance outcomes [38].

Morocco is a country that has just introduced integrated STEM education into its elementary science curriculum, following a revision of the elementary science curriculum in 2020. The purpose of introducing integrated STEM education is to promote the teaching of STEM subjects and enable Moroccan elementary school students to make sense of their learning in these subjects [39]. The STEM model adopted in the Moroccan elementary school curriculum focuses on engineering and technical design activities as contexts for students to apply and integrate their learning in science, technology, and mathematics to solve problems.

In order to improve learning in STEM subjects and increase the quality of STEM projects, the Moroccan elementary school curriculum recommends that teachers use information and communication technologies (ICT) in teaching and learning activities. Specifically, it is recommended that Moroccan teachers in elementary schools allow students to consult and use digital resources available through mobile devices (tablets and smartphones) inside or outside the classroom [39]. However, the innovations in the Moroccan elementary school curriculum, such as integrated STEM education and the use of mobile devices, cannot be effectively implemented in the classroom without in-service training programs for practicing teachers and pre-service training for future teachers. For these training programs to have a chance of success, they must take into account the recipients' beliefs and attitudes toward integrated STEM education and the use of mobile devices in STEM.

In an effort to improve the quality of primary teachers' initial training and promote their professionalization, Morocco launched a higher education program called the Bachelor of Education Elementary Specialty (BEES) at the beginning of the 2018-2019 academic year. Created as part of the 2015–2030 strategic vision launched by the Higher Council for Education, Training, and Scientific Research (HCETSR) in 2015, this program provides a long initial university education (3 years) leading to a Bachelor's degree in Education, which allows students to enter the Regional Centers for Careers in Education and Training (RCCET) through a competitive examination for a one-year professional training. This professional training leads to a certificate of pedagogical qualification for the elementary cycle, which allows its holder to work as a teacher in this cycle. The BEES allows students to benefit during their studies from training modules in languages, sciences, mathematics, educational sciences, ICT, and teaching methods of the different subjects taught in the elementary cycle [40].

In addition to providing future teachers with the theoretical, technical, and methodological foundations needed to teach in the elementary grades, the BEES training is expected to prepare these future teachers to implement the new pedagogical practices advocated by the new Moroccan elementary curriculum, such as integrated STEM education and the use of mobile devices in STEM education. The successful future implementation of these new practices depends heavily on the attitudes of these actors [41], [42]. Therefore, any improvement in these attitudes is likely to encourage future teachers to later implement integrated STEM education and the use of mobile devices to teach STEM in their classrooms [15], [16], [41], [43].

Although three years have passed since the implementation of the new Moroccan elementary education curriculum, almost nothing is known about the attitudes of pre-service elementary school teachers toward integrated STEM education and the use of mobile devices in this education. Likewise, little is known about the impact of pre-service training programs on the intentions of Moroccan elementary future teachers to implement integrated STEM education and to use mobile devices in that education. In this context, and to fill some of these knowledge gaps, this study focused on BEES students as pre-service teachers to examine their attitudes toward integrated STEM education and the use of mobile devices in education. Therefore, the current study attempts to provide sufficient answers to the following research questions: i) What are the attitudes of Moroccan BEES students toward integrated STEM education? Are there significant differences in their attitudes by gender, grade level, and pre-university studies? (RQ1); and ii) What are the attitudes of Moroccan BEES students toward the use of mobile devices in integrated STEM education? Are there significant differences in their attitudes by gender, grade level, and pre-university studies? (RQ2).

## 2. METHOD

This exploratory study focused on Moroccan BEES students and their attitudes toward the implementation of integrated STEM education in the Moroccan elementary school cycle. It also examined these students' attitudes toward the use of mobile devices in STEM projects, as advocated by the curriculum for this cycle. A questionnaire was used to collect responses from participants, and descriptive statistics (means and standard deviations) and Mann-Whitney tests were used to analyze the attitude results. The population of the study was made up of Moroccan students in their second- and third-year (final year) of BEES. The first-year teacher students were excluded from the survey because they did not receive any training on teaching methods for the different subjects taught in the elementary cycle in the first year, which could bias the results of the survey, especially in the part measuring these teacher candidates' self-confidence in implementing integrated STEM education. In contrast, the second- and third-year teacher candidates received all the planned BEES training modules on the subjects taught in the Moroccan elementary cycle and on the methodologies for teaching these subjects [40].

The data were collected through a questionnaire. The study group included a total of 226 students in BEES at the Higher Normal School (HNS) of Tetouan of Abdelmalek Essaadi University and the Higher School of Education and Training (HSET) of Ibn Tofail University of Kenitra in Morocco. The convenience sampling method was used to select the research sample. This is a non-probability sample in which the authors of this study selected individuals who were both readily available and willing to participate in the study [44]. Students were informed in advance about the objectives of the study, how their data would be used, and the measures taken to ensure the confidentiality and anonymity of the data collected. They were then invited to participate in the research on a voluntary basis.

The questionnaire was made available online through an online survey platform set up by the authors of this study, and all responses collected were anonymized to protect the confidentiality and privacy of study participants. The questionnaire consisted of 3 sections and 28 items (questions). The first section contained 6 questions that collect personal and demographic information about BEES students. The second section contained 13 items to measure attitudes toward integrated STEM education. The third section included 9 items to measure attitudes toward the use of mobile devices in integrated STEM education. A five-point Likert scale with strongly agree (5), agree (4), neutral (3), disagree (2), and strongly disagree (1) was used to measure the 22 attitude items toward integrated STEM education and the use of mobile devices in this education. Descriptive statistics (means and standard deviations) and Mann-Whitney tests were used to analyze the attitude results. The questionnaire was developed by the authors of this study. Some items used to measure BEES students' attitudes toward integrated STEM education were adopted from other studies [45], [46]. Some items used to measure BEES students' attitudes toward the use of mobile devices in integrated STEM education were adopted from several relevant studies [23], [33], [47], [48]. The questionnaire was submitted to two expert professors from Abdelmalek Essaadi University, Morocco, to check the clarity of the questions. After a thorough review, the experts concluded that the questions were appropriate for BEES students and addressed the intended research questions.

An exploratory factor analysis was conducted on items related to attitudes toward integrated STEM education and the use of mobile devices in integrated STEM education. The validity of the items was tested with a factor analysis conducted using the extraction method of principal component analysis with varimax rotation. Items measuring participants' attitudes toward integrated STEM education loaded into three factors: importance of integrated STEM education (between 0.752 and 0.926), enjoyment of implementing integrated STEM education (between 0.618 and 0.923), and self-confidence to implement integrated STEM education (between 0.707 and 0.835). Items measuring BEES students' attitudes toward the use of mobile devices in integrated STEM education were loaded as a single factor (between 0.614 and 0.864). Reliability tests were conducted for the different constructs by calculating Cronbach's alpha. The Cronbach's alpha values for the items in the different constructs were greater than 0.7, as shown in Table 1, which means that the items in these constructs had an acceptable level of internal consistency [49].

Table 1. Cronbach's alpha of the constructs

| Constructions and dimensions | Cronbach-α |
|---|---|
| Attitude toward integrated STEM education | 0.84 |
| Importance of integrated STEM education | 0.91 |
| Enjoyment of implementing integrated STEM education | 0.89 |
| Self-confidence to implement integrated STEM education | 0.78 |
| Attitudes toward the use of mobile devices in integrated STEM education | 0.91 |

The three constructs (dimensions) that grouped the items used to measure student attitudes toward integrated STEM education were used to assess the cognitive aspects (importance of STEM and self-confidence in implementing integrated STEM education) as well as the affective components (enjoyment of implementing integrated STEM education) of these attitudes. BEES students' attitudes toward integrated STEM education were measured using 13 items organized into three dimensions: "importance of integrated STEM education" (ISTEM) (measured by 4 items), "enjoyment of implementing integrated STEM education" (ESTEM) (measured by 5 items), and "self-confidence in implementing integrated STEM education" (SCSTEM) (measured by 4 items). The mean of each dimension was calculated by combining the items in that dimension, and the mean of attitudes toward STEM was calculated by combining the items in all three dimensions.

BEES students' attitudes toward the use of mobile devices in integrated STEM education were measured using 9 items. Of the 9 items used to measure each attitude, 3 items were used to measure perceived usefulness of mobile devices (PU), 2 items were used to measure perceived ease of use of mobile devices (PEU), 2 items were used to measure perceived enjoyment of using mobile devices (EU), and 2 items were used to measure intention to use mobile devices (IU). The mean of attitudes toward the use of mobile devices in integrated STEM education was calculated by combining the 9 items used to measure these attitudes.

Data analysis was performed using SPSS version 26. Means and standard deviations were used to analyze and interpret the attitude results. The scores used to interpret the attitude results were as follows: strongly disagree, 1.00-1.79; disagree, 1.80-2.59; have no idea, 2.60-3.39; agree, 3.40-4.19; strongly agree, 4.20-5.00. Mann-Whitney U tests were used to examine differences by gender, grade level, and pre-university studies between the mean scores of students' responses to items measuring different dimensions of attitudes toward integrated STEM education and the use of mobile devices in STEM education. The Mann-Whitney U test was preferred to the t-test because the data were not normally distributed. For each Mann-Whitney test, a difference was considered statistically significant if $p<0.05$. When significant differences were found by a Mann-Whitney test, the following intervals were used to define the effect size: $r<0.3$, small; $0.3 \leq r < 0.5$, medium; $r \geq 0.5$, large [50].

## 3. RESULTS AND DISCUSSION
### 3.1. Demographic data of the sample

Table 2 shows that the majority of students who participated in the survey were female, in their third year of BEES, were between the ages of 19 and 21 (mean age =20.38 years, SD=1.04), and all participants owned a mobile device. It is important to note that students who have access to BEES may have different pre-university educational backgrounds, including scientific, literary, or technological baccalaureate degrees. The diversity of the pre-university educational backgrounds of BEES students means that these students do not benefit equally from certain STEM subjects or the time envelopes of certain STEM subjects. A scientific baccalaureate allows students to benefit from larger time envelopes in mathematics and science compared to students with a literary baccalaureate. Some science pre-university backgrounds also allow students to benefit from technology and engineering courses or workshops, while other backgrounds do not allow students to benefit from such courses and workshops. Table 2 shows that the majority of students who participated in the survey had a scientific baccalaureate degree.

Table 2. Survey participant demographics

|  |  | n | Percentages (%) |
|---|---|---|---|
| Gender | Female | 172 | 76.11 |
|  | Male | 54 | 23.89 |
| Age | 18 | 3 | 1.33 |
|  | 19 | 46 | 20.35 |
|  | 20 | 72 | 31.86 |
|  | 21 | 79 | 34.96 |
|  | 22 | 19 | 8.41 |
|  | 23 | 7 | 3.10 |
| Grade level | Second year of BEES | 81 | 35.84 |
|  | Third year of BEES | 145 | 64.16 |
| Type of baccalaureate | Scientific baccalaureate | 174 | 76.99 |
|  | Literary baccalaureate | 52 | 23.01 |
| Mobile device ownership |  | 226 | 100 |

### 3.2. Research question 1
#### 3.2.1. Attitudes of Moroccan BEES students toward integrated STEM education

In analyzing the attitudes of Moroccan BEES students toward integrated STEM education, this study found that the students' responses to the items used in this study to measure these attitudes were generally neutral, meaning that these students' attitudes toward integrated STEM education were moderate as shown in Table 3. Participants' responses to items on the "importance of integrated STEM education" dimension were generally centered around the "agree" option, and responses to items on the "enjoyment of implementing integrated STEM education" dimension were generally centered around the "no idea" option. In contrast, responses to items on the "confidence in implementing integrated STEM education" dimension were generally centered around the "disagree" option. Therefore, it can be inferred that BEES students believe that integrated STEM education is important, they are not sure that they will enjoy this education, and they are not confident in their ability to implement it. This result is consistent with the findings of Abdullah *et al.* [51] who reported high cognitive and moderate affective teacher readiness to implement integrated STEM education. This result is also consistent with the findings of the systematic review by Margot and Kettler [8], which showed that teachers are aware of the importance of STEM but are not confident in their ability to implement integrated STEM education. In addition, this finding is in line with other previous studies that reported teachers' unpreparedness to teach STEM [52] and lack of confidence in their ability to teach STEM in the classroom [53].

The present study's outcomes reveal substantial deficiencies in pedagogical training for BEES students' integrated STEM learning. In this context, the participants' responses to items in the 'self-confidence in implementing integrated STEM education' dimension indicate that they lacked sufficient pedagogical knowledge about integrated STEM education, felt inadequately prepared to implement it in an elementary classroom, were unable to answer elementary students' questions about STEM projects, and did not know how to assist these students in succeeding with their STEM projects. These findings are consistent with prior research, including the studies conducted by Margot and Kettler [8] and Pimthong and Williams [54], which uncovered inadequacies in integrated STEM pre-service teacher training. Similarly, these findings are consistent with the study by Susanti *et al.* [55] which reported that the majority of primary school teachers in one region of Indonesia had knowledge gaps in STEM and were inadequately prepared to teach in this area. In addition, these findings are in line with the findings of Kurup *et al.* [56] which indicated that pre-service elementary school teachers had limited understanding of STEM and limited confidence in teaching STEM due to their limited experience in teaching STEM during their university training and professional practice. Therefore, it can be concluded that BEES students' lack of knowledge about integrated STEM education affects their preparation [10] and may explain their lack of confidence in their ability to implement integrated STEM projects.

Table 3. Means and standard deviations of items and dimensions of attitude toward integrated STEM education

| Constructions and items | | N | M | SD | Interpretation |
|---|---|---|---|---|---|
| Attitudes toward integrated STEM education. | | 226 | 2.92 | 0.40 | No idea |
| ISTEM | Importance of integrated STEM education | 226 | 4.15 | 0.57 | Agree |
| ISTEM1 | I think that integrated STEM education will help elementary school students in their future work. | 226 | 4.17 | 0.68 | Agree |
| ISTEM2 | I think that integrated STEM education will help elementary students throughout their school careers. | 226 | 4.16 | 0.62 | Agree |
| ISTEM3 | Solving real-world problems in integrated STEM education increases elementary school students' interest in technology and engineering. | 226 | 4.08 | 0.66 | Agree |
| ISTEM4 | Solving real-world problems in integrated STEM education increases elementary school students' interest in science and mathematics. | 226 | 4.18 | 0.59 | Agree |
| ESTEM | Enjoyment of implementing integrated STEM education | 226 | 2.61 | 0.60 | No idea |
| ESTEM1 | I am interested in integrated STEM education at the elementary school. | 226 | 2.35 | 0.75 | Disagree |
| ESTEM2 | As a teacher, I would like to engage elementary school students in STEM projects. | 226 | 2.36 | 0.73 | Disagree |
| ESTEM3 | I would like to participate in training programs that help elementary school teachers implement STEM projects. | 226 | 3.36 | 0.65 | Agree |
| ESTEM4 | Primary students enjoy participating in STEM projects. | 226 | 2.77 | 0.76 | No idea |
| ESTEM5 | As a teacher, I am sure I will be able to get elementary school students to appreciate STEM projects. | 226 | 2.23 | 0.69 | Disagree |
| SCSTEM | Self-confidence to implement integrated STEM education | 226 | 2.08 | 0.54 | Disagree |
| SCSTEM1 | I have sufficient pedagogical knowledge about integrated STEM education. | 226 | 2.28 | 0.66 | Disagree |
| SCSTEM2 | I feel sufficiently prepared to implement integrated STEM education in an elementary school classroom. | 226 | 2.03 | 0.76 | Disagree |
| SCSTEM3 | In general, I think I can answer questions from elementary school students about STEM projects. | 226 | 2.02 | 0.59 | Disagree |
| SCSTEM4 | I know how to help elementary school students see their STEM projects through to completion. | 226 | 1.99 | 0.74 | Disagree |

### 3.2.2. Differences in Moroccan BEES students' attitudes toward integrated STEM education by gender, grade level, and pre-university studies

The Mann-Whitney U test used to examine the relationship between the gender of BEES students and their attitudes toward integrated STEM education, as shown in Table 4, revealed no significant differences between the mean scores of male and female students' responses to items measuring the dimensions: "importance of integrated STEM education," "enjoyment of implementing integrated STEM education," and "self-confidence in implementing integrated STEM education." This study also found that BEES students' attitudes toward integrated STEM education do not differ significantly by gender. This result corroborates other findings in the literature, which also indicated that gender does not affect pre-service teachers' attitudes toward integrated STEM education [11], [57]–[61]. Similarly, previous studies [53], [62] have found no significant relationship between elementary and secondary teachers' gender and their attitudes toward integrated STEM education. The lack of gender effect on BEES students' attitudes toward integrated STEM education could be due to the fact that these students are in the same university training and have undergone almost similar pre-university studies without gender differentiation.

Four Mann-Whitney U tests were conducted, as shown in Table 5, to examine the relationship between BEES students' grade level and their attitudes toward integrated STEM education. The first test indicated no statistically significant difference between the mean scores of the second- and third-year students' responses to all of the items used to measure attitudes toward integrated STEM education. The other three tests also indicated no significant differences between the mean scores of students' responses at the two levels to items measuring the dimensions of "importance of integrated STEM education," "enjoyment of implementing integrated STEM education," and "self-confidence to implement integrated STEM education."

Based on the previous tests, it can be concluded that there is no significant relationship between students' grade level and their attitudes toward integrated STEM education. In addition, there is no significant relationship between students' grade level and their responses to questions measuring the three dimensions of attitudes toward integrated STEM education. This finding is consistent with the results of the study by Hacıömeroğlu [58], who found no significant differences in pre-service elementary teachers' knowledge, subjective norms, and attitudes toward integrated STEM education based on their grade level. Similarly, Temel [61] reported no significant relationship between pre-service elementary teachers' grade level and their attitudes toward STEM education. This finding is also consistent with the result of the study by Ateş and Gül [63], who found no significant relationship between pre-service teachers' educational level and their self-efficacy beliefs about STEM education. In contrast, this finding contradicts the result of the study by Kartal and Taşdemir [11], who showed that pre-service teachers' attitudes toward STEM tended to be more positive at higher levels of study than at lower levels because the knowledge and skills of the participants in this study improved as they progressed through the training programs. In the present study, the absence of an impact of BEES students' grade level on their attitudes toward integrated STEM education can be attributed to the fact that students do not receive any courses or workshops on integrated STEM education during their three years of training, which could potentially develop their attitudes toward this education.

The diversity of BEES students' pre-university backgrounds necessitates some interest in looking for possible relationships between these backgrounds and these students' attitudes toward integrated STEM education. As shown in Table 6, the Mann-Whitney U test used to compare the attitudes of BEES students with a scientific baccalaureate degree and those with a literary baccalaureate degree toward integrated STEM education revealed statistically significant differences between the means of the responses of these two groups of students to the items used to measure attitudes toward integrated STEM education. The mean response scores for students with a scientific baccalaureate degree are higher than those for students with a literary baccalaureate degree. The differences in attitudes toward integrated STEM education between scientific and literary baccalaureate students are moderate [50]. Furthermore, the other three Mann-Whitney U tests in Table 6, which were used to examine the differences between the mean scores of the responses of BEES students with a scientific baccalaureate and those of students with a literary baccalaureate to the items used to measure "importance of integrated STEM education," "enjoyment of implementing integrated STEM education," and "self-confidence in implementing integrated STEM education," revealed significant differences between the responses of the two groups. The mean response scores of students with a scientific bachelor's degree are higher than those of students with a literary bachelor's degree. The differences between the mean scores of the responses of the two groups of students to the items measuring "importance of integrated STEM education" and "self-confidence in implementing integrated STEM education" are moderate, while the differences between the mean scores of the responses of the two groups of students to the items measuring "enjoyment of implementing integrated STEM education" are small [50].

Table 6 results indicate that students with a scientific baccalaureate generally have more positive attitudes toward integrated STEM education than students with a literary baccalaureate. In addition, BEES students with a scientific baccalaureate are generally more aware of the importance of integrated STEM education and more confident in their ability to implement it than their peers with a literary baccalaureate. This trend can be explained by the higher number of science and mathematics courses taken by students with a scientific baccalaureate during their pre-university studies compared to those with a literary baccalaureate. This increased exposure to STEM subjects enables them to acquire more knowledge in these fields, which could explain their more positive attitudes towards integrated STEM education. This finding is consistent with the study by Nadelson et al. [64] which found a positive correlation between teachers' knowledge of STEM subjects and their confidence in teaching STEM. Teachers with more knowledge in these subjects are generally more confident in their ability to teach STEM effectively [65]. Similarly, Margot and Kettler [8] found that teachers are more comfortable teaching STEM when they have taken more STEM-related courses. The results of this study are also supported by previous research [11], [66], which found significant differences in pre-service teachers' attitudes toward STEM in favor of those specializing in science.

Table 4. Results of Mann-Whitney tests examining relationships between gender and BEES students' attitudes toward integrated STEM education

| Attitudes and dimensions | BEES female students (n=172) | | | BEES male students (n=54) | | | Mann-Whitney U test | | |
|---|---|---|---|---|---|---|---|---|---|
| | M | SD | Mdn | M | SD | Mdn | U | z | p |
| Attitudes toward integrated STEM education. | 2.92 | 0.42 | 2.96 | 2.91 | 0.37 | 2.94 | 4502.5 | -0.338 | 0.735 |
| Importance of integrated STEM education | 4.17 | 0.57 | 4.07 | 4.09 | 0.55 | 4.02 | 4288 | -0.887 | 0.375 |
| Enjoyment of implementing integrated STEM education | 2.60 | 0.60 | 2.62 | 2.66 | 0.61 | 2.82 | 4335 | -0.743 | 0.457 |
| Self-confidence to implement integrated STEM education | 2.09 | 0.56 | 2.11 | 2.06 | 0.46 | 2.03 | 4385.5 | -0.624 | 0.533 |

Table 5. Results of Mann-Whitney tests examining the relationships between BEES students' educational level and their attitudes toward integrated STEM education

| Attitudes and dimensions | Second year students (n=81) | | | Third year students (n=145) | | | Mann-Whitney U test | | |
|---|---|---|---|---|---|---|---|---|---|
| | M | SD | Mdn | M | SD | Mdn | U | z | p |
| Attitudes toward integrated STEM education. | 2.89 | 0.38 | 2.94 | 2.93 | 0.42 | 2.96 | 5416 | -0.971 | 0.332 |
| Importance of integrated STEM education | 4.11 | 0.55 | 4.07 | 4.17 | 0.58 | 4.05 | 5844 | -0.063 | 0.950 |
| Enjoyment of implementing integrated STEM education | 2.63 | 0.54 | 2.62 | 2.61 | 0.64 | 2.71 | 5798 | -0.159 | 0.873 |
| Self-confidence to implement integrated STEM education | 1.99 | 0.49 | 2.04 | 2.13 | 0.56 | 2.14 | 5088 | -1.683 | 0.092 |

Table 6. Results of Mann-Whitney tests examining the relationships between pre-university studies and BEES students' attitudes toward integrated STEM education

| Attitude and dimensions | Students with a scientific baccalaureate degree (n=174) | | | Students with a literary baccalaureate degree (n=52) | | | Mann-Whitney U test | | | |
|---|---|---|---|---|---|---|---|---|---|---|
| | M | SD | Mdn | M | SD | Mdn | U | z | p | r |
| Attitudes toward integrated STEM education. | 3.01 | 0.38 | 3.03 | 2.62 | 0.34 | 2.50 | 2009.5 | -6.091 | 0.000* | 0.39[b] |
| Importance of integrated STEM education | 4.24 | 0.58 | 4.14 | 3.83 | 0.38 | 3.84 | 2214.5 | -5.831 | 0.000* | 0.37[b] |
| Enjoyment of implementing integrated STEM education | 2.65 | 0.62 | 2.80 | 2.50 | 0.54 | 2.38 | 3649.5 | -2.131 | 0.033** | 0.14[c] |
| Self-confidence to implement integrated STEM education | 2.24 | 0.44 | 2.23 | 1.56 | 0.49 | 1.60 | 1501.5 | -7.387 | 0.000* | 0.47[b] |

Note. [a] large effect size, [b] medium effect size, [c] small effect size, *p<0.001, **p<0.05

### 3.3. Research question 2
#### 3.3.1. Attitudes of Moroccan BEES students toward the use of mobile devices in integrated STEM education

The analysis of BEES students' attitudes toward the use of mobile devices in integrated STEM education as seen in Table 7 revealed that these attitudes are generally positive, as the mean scores of the items used to measure these attitudes were centered on the "agree" option. Additionally, BEES students perceive the use of mobile devices as beneficial to the practice of integrated STEM education, motivating elementary students to engage in STEM projects, and improving the quality of these projects. BEES students also expressed comfort with the use of mobile devices in integrated STEM education and their intention to incorporate this technology into their teaching. However, the mean score of the responses to the PEU2 item was generally centered on the "no idea" option, indicating that these BEES students are uncertain about the ability of elementary students to use mobile devices in STEM projects. The findings of this study regarding the attitudes of BEES students as future teachers toward the use of mobile devices in integrated STEM education are consistent with the findings of other studies in the literature that have reported positive attitudes of pre-service teachers toward the use of mobile devices in their future teaching practice [67]–[71].

This study found that BEES students have positive attitudes toward the use of mobile devices in integrated STEM education, even though they previously reported that they do not have enough pedagogical knowledge about integrated STEM education, do not feel prepared to implement integrated STEM education, are not able to answer students' questions about their STEM projects, and do not know how to help elementary students complete their STEM projects. BEES students' positive attitude toward the use of mobile devices in integrated STEM education, despite their reported lack of pedagogical knowledge about such education, indicates that these students view technological knowledge as distinct from pedagogical knowledge [72], which would enable them, as teachers, to effectively use mobile technology in integrated STEM education despite their lack of adequate pedagogical knowledge about implementing such education. This finding may be explained by the fact that in the first year of the BEES program, students receive general training modules in the use of ICT, and only in the second year do they receive training in pedagogical strategies for teaching different subjects. This sequence of training modules may not allow BEES students to establish connections between technological tools and the pedagogical strategies of the subjects in which these tools may be used.

Table 7. Means and standard deviations of items and attitudes toward the use of mobile devices in integrated STEM education

| Constructions and items | | N | M | SD | Interpretation |
|---|---|---|---|---|---|
| Attitudes toward the use of mobile devices in integrated STEM education | | 226 | 3.91 | 0.63 | Agree |
| PU1 | The use of mobile devices (tablets and smartphones) has a positive impact on integrated STEM education practices. | 226 | 3.78 | 0.88 | Agree |
| PU2 | Primary school students will be more motivated to engage in STEM projects if they use mobile devices (tablets and smartphones). | 226 | 4.19 | 0.72 | Agree |
| PU3 | The use of mobile devices (tablets and smartphones) improves the ability of elementary school students to work on STEM projects. | 226 | 3.80 | 0.92 | Agree |
| PEU1 | As an elementary school teacher, I will have no problem using mobile devices (tablets and smartphones) to mentor students' STEM projects. | 226 | 4.23 | 0.60 | Strongly agree |
| PEU2 | Primary school students have no problem using mobile devices (tablets and smartphones) to carry out their STEM projects. | 226 | 2.94 | 1.00 | No idea |
| EU1 | As an elementary school teacher, I would like to use mobile devices to mentor students on STEM projects. | 226 | 3.92 | 0.86 | Agree |
| EU2 | Primary school students like to use mobile devices (tablets and smartphones) when working on their STEM projects. | 226 | 3.96 | 0.81 | Agree |
| IU1 | As an elementary school teacher, if my school has mobile devices (tablets and smartphones), I will have students use these devices to develop their STEM projects. | 226 | 4.27 | 0.77 | Strongly agree |
| IU2 | If an elementary school has mobile devices (tablets and smartphones), I recommend that teachers at that school use these mobile devices to help their students develop STEM projects. | 226 | 4.15 | 0.79 | Agree |

### 3.3.2. Differences in Moroccan BEES students' attitudes toward the use of mobile devices in integrated STEM education by gender, grade level, and pre-university studies

Mann-Whitney U tests were used to examine differences in the attitudes of BEES students by gender, grade level, and pre-university studies toward the use of mobile devices in integrated STEM education. The test examining the relationship between the gender of BEES students and their attitudes toward the use of mobile devices in integrated STEM education revealed no statistically significant difference in attitudes between male and female BEES students (U=4442.5, Z=-0.484, p=0.628). Thus, it can be deduced that there is no relationship between the gender of BEES students and their attitudes toward the use of mobile devices in integrated STEM education. This finding corroborates other studies that have found no significant gender differences in preservice teachers' attitudes toward the use of mobile devices [68], [73]–[75]. The lack of effect of BEES students' gender on their attitudes toward the use of mobile devices in integrated STEM education may be due to the fact that all students own mobile devices, which has allowed them to become familiar with the use of these devices, and that they all received the same training in the use of ICT.

The test used to examine the relationship between BEES students' grade level and their attitudes toward the use of mobile devices in integrated STEM education showed no statistically significant differences in the attitudes of second- and third-year students (U=5324, Z=-1.169, p=0.242). Therefore, it can be concluded that there is no relationship between the grade level of the BEES students and their attitudes toward the use of mobile devices in integrated STEM education. This result is consistent with the findings of other studies that found no significant differences in pre-service primary teachers' attitudes toward the use of mobile technologies based on their grade level [69], [70], [76]. The results of the present study can be attributed to the fact that all participants received two identical modules of information and communication technology training in the first year of the bachelor's degree. In addition, the third-year students did not receive additional training specific to the use of mobile devices in their future teaching practice compared to their second-year peers.

The Mann-Whitney test used to examine the relationship between students' pre-university studies and their attitudes toward the use of mobile devices in integrated STEM education revealed no significant difference in attitudes between students with a scientific baccalaureate and those with a literary baccalaureate (U=4256.5, Z=-0.650, p=0.516). Thus, it can be concluded that there is no relationship between BEES students' pre-university studies and their attitudes toward the use of mobile devices in integrated STEM education. This result may be due to the fact that none of the pre-university courses offer Moroccan students the opportunity to use mobile devices in their learning activities and the fact that all the students who participated in this survey own mobile devices, which would have given them sufficient experience in using these devices and, consequently, would have strengthened their positive attitudes toward the use of mobile devices in their future teaching practices [77].

The results of this study have several notable implications for the training of primary school teachers in Morocco to implement integrated STEM education and the use of mobile devices in this context. First, they highlight the imperative need to provide BEES students with theoretical courses on the content and pedagogical strategies of integrated STEM education, as well as practical workshops. This could improve their knowledge of integrated STEM education and increase their confidence in their ability to implement STEM projects in elementary schools [8], [78]. Second, given the significant differences in BEES students' attitudes toward integrated STEM education based on their pre-university studies, the study suggests that the BEES program should allow students with a literary baccalaureate to take more courses in STEM subjects. This is to close the gap created by their previous studies, which did not allow them to benefit from as many STEM courses as their peers with a scientific baccalaureate. Third, to prevent future teachers' uncertainty about elementary students' ability to use mobile devices from hindering the implementation of this technology, this study recommends that BEES training should prepare pre-service teachers to overcome the challenges that elementary students may face. Fourth, because BEES students do not perceive the relationship between content knowledge, pedagogical knowledge, and technological knowledge, this study recommends that BEES training should enable them to understand the interdependence of these three areas for the use of mobile devices in integrated STEM education through theoretical courses and practical workshops [72]. Finally, the lack of a significant effect of both gender and pre-university studies on attitudes toward mobile learning suggests that all BEES students could benefit from the previously recommended mobile learning training without the need to differentiate based on these two factors.

## 4. CONCLUSION

This study contributes to the understanding of pre-service elementary teachers' attitudes towards integrated STEM education and the use of mobile devices in Morocco. Although they recognize the importance of STEM, these pre-service teachers express a lack of confidence in their ability to provide effective STEM education. These concerns highlight the need to strengthen pre-service teacher education in the theoretical and practical areas of integrated STEM education. Notably, the study reveals significant differences in pre-service teachers' attitudes toward integrated STEM education based on their pre-university studies. This finding suggests the need to adapt the current BEES curriculum to address educational inequalities resulting from different pre-university pathways that do not provide equal access to STEM courses. In addition, although pre-service teachers are generally in favor of integrating mobile devices into STEM education, they have reservations about elementary students' ability to use these devices effectively in STEM projects and do not perceive the interrelationship between content knowledge, pedagogical strategies, and technological tools in the context of implementing mobile technology. These observations point to the need for targeted educational interventions to improve pre-service teachers' technical and pedagogical skills in using mobile technologies in STEM education. However, it is important to consider the limitations of this study. The data are self-reported and collected through an online questionnaire, which may introduce bias. Furthermore, the study is based on a sample limited to two Moroccan universities, which limits its generalizability. Therefore, future research should seek to replicate this study in several Moroccan universities with larger samples and using a variety of data collection methods such as interviews and focus groups to complement questionnaires.

## BIOGRAPHIES OF AUTHORS

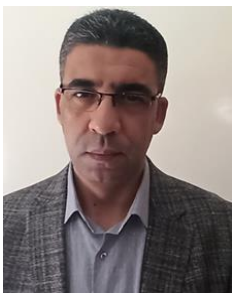

**Aziz Amaaz** has worked as a secondary school technology teacher and is currently a technology educational inspector. He is currently pursuing a Ph.D. in physics education at the Faculty of Science, Abdelmalek Essaadi University. He is a member of the "Computer Science and University Pedagogical Engineering" research team. His research interests include physics education, integrated STEM education, the integration of information and communication technologies in STEM education, and the design and development of digital resources for mobile learning. He can be contacted at email: aamaaz@uae.ac.ma.

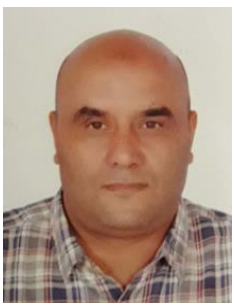

**Abderrahman Mouradi** has worked as an educational technology inspector. He is currently a professor of higher education in physics and science didactics at the Higher Normal School of Abdelmalek Essaâdi University, Morocco. He teaches physics and science didactic modules for future physics and primary school teachers. He also conducts research on wind energy and physics education, with a focus on STEM, the integration of new technologies in physics teaching and learning, and the use of augmented reality in physics education. He is active in editing and publishing scientific journals and has currently published 7 articles in indexed journals and presented over 15 papers at international conferences. He can be contacted at email: abderrahman.mouradi@uae.ac.ma.

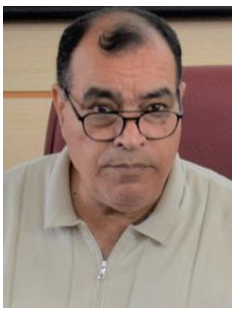

**Moahamed Erradi** is a professor at the Higher Normal School of Abdelmalek Essaadi University, Morocco. He teaches science didactics and multimedia pedagogical engineering. He is the director of the "Computer Science and University Pedagogical Engineering" research team at Higher Normal School of Abdelmalek Essaadi University and a member of the "Laboratory of Operational Research, Computing, and Applied Statistics" (LIROSA) at the Faculty of Science, Abdelmalek Essaadi University. He is also the Head of the Department of Mathematics, Physics and Computer Science at the Higher Normal School, Abdelmalek Essaadi University. His research interests include physics education, science didactics, multimedia pedagogical engineering, integration of information and communication technologies in science teaching and learning, and adaptive learning. He can be contacted at email: m.erradi@uae.ac.ma.

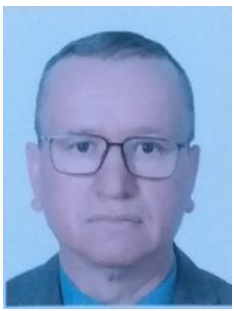

**Ali Allouch** is professor of higher education in physical sciences and didactics of physics and chemistry at the Higher School of Education and Training, Ibn Tofail University, Kenitra, Morocco. This school is dedicated to the training of primary and secondary school teachers in Morocco. He is working on an innovative design project in collaboration with the "Mechanical and Integrated Engineering Team M2I" laboratory at the National Higher School of Arts and Crafts (ENSAM), Moulay Ismail University, Meknes, Morocco. His research also focuses on physics education and integrated STEM education. He has participated as a reviewer at several international STEM conferences. He is working as a USAID expert on redesigning science curricula with a STEM approach at the Ministry of National Education in Morocco. He can be contacted at email: ali.allouch@uit.ac.ma.